\pdfoutput=1
\documentclass[runningheads]{llncs}

\usepackage[T1]{fontenc}
\usepackage{amssymb}
\usepackage{graphicx}
\usepackage{color}
\usepackage{xcolor}
\usepackage{hyperref}
\hypersetup{
    colorlinks=true,
    linkcolor=teal,
    filecolor=purple,      
    urlcolor=teal
    }

\newcommand{\blue}[1]{#1}

\newcommand{\ul}[1]{\underline{#1}}
\newcommand{\uul}[1]{\underline{\underline{#1}}}
\newcommand{\alp}{^{(\alpha)}}

\newcommand{\Ta}{ {\uul{T}\alp}}
\newcommand{\Ma}{ {\uul{M}\alp}}

\newcommand{\Tk}{\mathrm{T}_s}

\begin{document}

\title{Parking search in urban street networks: Taming down the complexity of the search-time problem via a coarse-graining approach}

\titlerunning{Taming down the complexity of parking search calculations}

\author{Léo BULCKAEN\inst{1,2} \and
Nilankur DUTTA\inst{1}\orcidID{0000-0002-6438-4985} \and
Alexandre NICOLAS\inst{1}\orcidID{0000-0002-8953-3924}}
\authorrunning{L. Bulckaen et al.}
%
\institute{$^1$ Institut Lumière Matière, CNRS and Université Claude Bernard Lyon 1, \\
         F-69622 Villeurbanne,         France\\
         \email{leo.bulckaen@polytechnique.edu}, 
\email{alexandre.nicolas@polytechnique.edu} \\
\url{} 
$^2$ Ecole polytechnique,
        F-91128, Palaiseau, France
}

\maketitle              
\begin{abstract}
The parking issue is central in transport policies and drivers' concerns, but the determinants of the parking search time remain relatively poorly understood. The question is often handled in a fairly \emph{ad hoc} way, or by resorting to crude approximations. Very recently, we proposed a more general agent-based approach, which notably takes due account of the role of the street network and the unequal attractiveness of parking spaces, and showed that it can be solved analytically by leveraging the machinery of Statistical Physics and Graph Theory, in the steady-state mean-field regime. Although the analytical formula is computationally more efficient than direct agent-based simulations,  it involves cumbersome matrices, with linear size proportional to the number of parking spaces. Here, we extend the theoretical approach and demonstrate that it can be further simplified, by coarse-graining the parking spot occupancy at the street level. This results in even more efficient analytical formulae for the parking search time, which could be used efficiently by transport engineers.


\end{abstract}

\date{}

\section{Introduction}

Parking is a complex problem of great practical as well as theoretical interest. On the theoretical side, 
its complexity arises from the interaction between multiple entities (cars) which have different destinations and parking preferences, as well as several possible states (driving, searching for parking, or parked), and whose motion is constrained by the network of streets: this complexity would obviously vanish into thin air if one were to consider a predictable single driver trying to park in an empty city. The problem thus presents a singular interplay between facets including collective effects, complex networks,  psychological factors, impact of transport policies.

On the practical side, the quandary of parking search is all too well known to individual drivers as well as transport authorities in virtually all large metropolitan areas \cite{shoup2018parking}. Motorists may spend several dozens of hours every year searching for parking, according to INRIX survey data \cite{INRIX2017}, whereas the former increasingly regard parking as a
lever to enforce their transport policy. It has been assessed that cars cruising for parking may represent a significant share of the total traffic 
in many large cities (e.g., 15\% in central Stuttgart, 28\% to 45\% in New York) \cite{shoup2018parking,hampshire2018share} and \blue{aggravate} congestion and pollution in city centres.

A deeper understanding of the process of parking search is thus crucial, so as to be able to predict the impact of hypothetical measures. Very recently, we put forward a general framework which goes beyond conventional numerical and theoretical approaches to parking search and which notably suitably accounts for the role of the street network and the unequal attractiveness of parking spaces \cite{dutta2022parking}. One major asset of this framework is that,
despite its generality, it permits analytic progress. Indeed, the problem can be solved not only by means of a computationally efficient agent-based algorithm that we developed, but also
by leveraging the powerful machinery of Statistical Physics and Graph Theory to obtain analytical formulae relating the search time and the occupancy of parking spots. While we showed  \cite{dutta2021searching,dutta2022parking} that this approach can be
applied to complex, large-scale networks such
as that of the city of Lyon, France, the analytical formulae were fairly impractical because they
involved cumbersome (even though sparse) matrices representing the graph of parking spots in the city. In this contribution, we purport to show that the complexity of the problem can be tamed down even more by considering a coarse-grained graph, whose smallest elements are street portions instead of individual parking spots. 

\section{Modelling framework} 

\subsection{Short review of existing agent-based approaches}

To start with, we very succinctly review some previous models developed to predict parking search times. The most basic model is probably
the \emph{binomial approximation} \cite{axhausen1994effectiveness}, which expresses the search time as $T_s \simeq \frac{T_0}{1-{\phi}}$, where $T_0$ is the time to drive from one spot to the next one. Unfortunately, this expression seems to strongly
underestimate search times, if it is used in conjunction with the reported occupancy $\phi$ in city districts; it can hardly be reconciled with the empirical observation of surging search times long before ${\phi}$ reaches 100\% \cite{belloche2015street,arnott2017cruising,gu2020macroscopic,weinberger2020parking}. 

To get insight 
into this mismatch, Arnott et al. \cite{arnott2017cruising} considered a simple model in which  cars moved along a circle with 100 spots along its contour and parked in the first available spot; they found that spatial and temporal correlations in the occupation of spots, among other factors, underlay the failure of the binomial approximation. 
\blue{
Belonging to the same kind of approaches revolving around simplified
networks, aimed at gaining general insight into the problem of parking search, Krapivsky and Redner
described the optimal parking strategy on a lane of parking spaces through the lens of statistical physics \cite{krapivsky2020should}, while Dowling and colleagues
analysed parking search in a regular network using the theory of network of finite-capacity queues
\cite{dowling2019modeling}.
} 

Aiming for a more detailed description, Levy et al. \cite{levy2015spatially} put forward the PARKAGENT model, more suitable for practical use, in which cars drive towards their destination on a spatially described network of streets and decide to park or not when driving by a vacant spot by estimating their odds to find another vacant spot closer to their target, on the basis of the occupancy of spots that they have seen so far. Should they reach their destination without having parked before, they will start circling and accept the first vacant spot and, after a fixed time threshold, they will drive to an off-street parking lot. 
Vo et al. \cite{vo2016micro} designed a simple, easy-to-use model in NETLOGO to predict the car movements in a parking lot. The model is based on a decision tree, which considers factors such as the existence of a vacant spot near the ticket machine or the entrance and the gender of the driver.

\blue{
Game-theoretic approaches have also been employed to address this problem, by supposing that a Nash equilibrium is reached by drivers intent on finding the best spot within a given, agent-specific search time, provided the reaching times for every parking space are known \cite{calise2022parking}.
}

\subsection{Presentation of the model}
The model that we recently introduced \cite{dutta2022parking} can be regarded as a general framework encompassing many of these agent-based models, insofar as drivers also move on a spatially described network of streets, with parking spots located along the streets, but their turn-choices and parking decisions can be prescribed arbitrarily.

\begin{figure}[!htb]
    \centering
        \includegraphics[width=\textwidth]{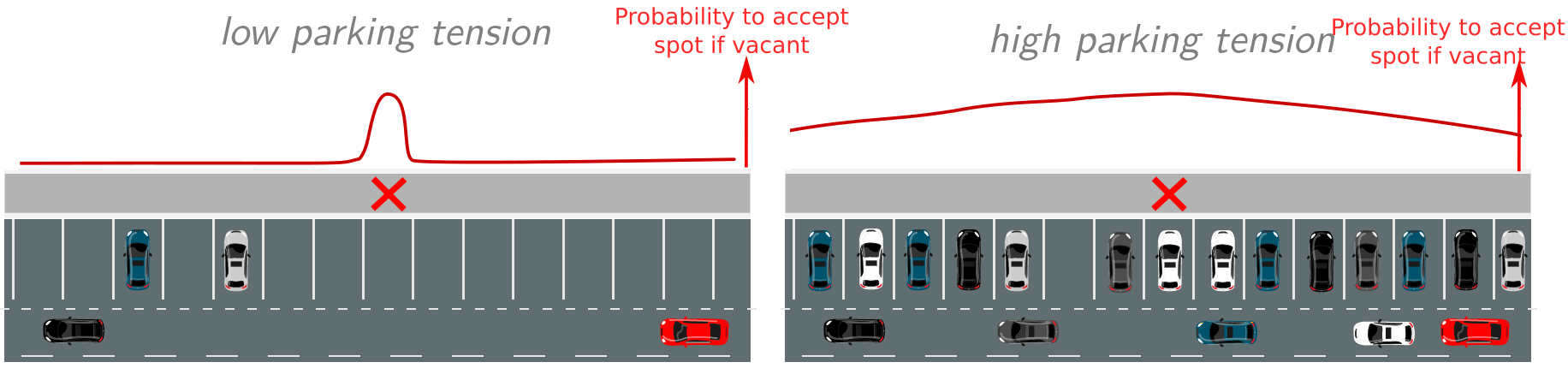}
\caption{Illustration of the effect of the parking tension on the probability to
accept to park at a vacant spot. On the left, parking tension is low, corresponding to $\beta \to \infty$, and
drivers will only accept to park at their favourite spots.}
\label{fig:parking_tension}
\end{figure}

More precisely,  
several categories $\alpha=1,\,2,\,\ldots$  of drivers can be defined depending on their destination, trip purpose, etc., and, at an intersection, drivers of each category have different probabilities to turn into 
the possible outgoing street links; these are given by the corresponding entry of a (category-dependent) turn-choice matrix ${\uul{T}}\alp$. Thus, each category of drivers may be routed to a different destination.

Besides, drivers of distinct categories will naturally differ in their decision to park or not when driving by
a vacant spot: they will choose to 
park there (if it is vacant) with probability $p_i\alp$, which in practice depends on a variety of explanatory variables, first of which how far it is from the destination, how much it costs, but also what are the odds of finding a `better' spot, e.g., closer to the target \cite{levy2013exploring,bonsall2004modelling}. To avoid prescribing specific rules for these parking choices, which are likely to depend on the local context, we chose to subsume all these factors into two generic variables, which can be tuned arbitrarily: (i) an attractiveness $A_i\alp$ reflecting how attractive a spot $i$ is perceived to be \emph{intrinsically}, (ii) the driver's perception of how easy it currently is to park, $\beta\alp \in [0,\infty)$. 
\begin{equation}
    p_i\alp(t)= f(A_i\alp,\beta\alp(t)).
\end{equation}
For simplicity, at present, the parameter $\beta\alp(t)$ measuring parking tension will always be a function of the \emph{global} occupancy ${\phi}(t)$, i.e., $\beta\alp(t) = f[\phi(t)] \equiv \beta$, even though more realistic dependencies could readily be contemplated.

\blue{As illustrated in Fig.~\ref{fig:parking_tension}, when the occupancy is very low,} parking seems extremely easy, which implies that $\beta\to \infty$, and the driver will refuse to park anywhere but in their preferred spot, of attractiveness $A_{\max}$. 
To the opposite, when the occupancy is very high, $\beta$ will tend to zero and the driver will accept virtually any admissible spot 
(of perceived attractiveness $A_i\alp>-\infty$), viz. $p_i\alp=1$.
Since $p_i\alp\in[0,1]$, these extreme cases are conducive to expressing $p_i$ with a Boltzmann-like functional form, viz.,
\begin{equation}
    p_i\alp= e^{\beta \cdot (A_i\alp-A_{\max})}.
\end{equation}

Finally, parked cars leave their space at a rate $D\alp$, which is the reciprocal of the  average parking duration. These departing cars are removed from the simulation, because the interaction
between cruising cars and the rest of the traffic is discarded here: cars move at fixed speeds in each street.

Thus formulated, our model offers a generalisation of existing agent-based approaches. For instance, if they prescribe to park in the first vacant spot within a radius of the destination \cite{fulman2021approximation}, this can be encoded in the model as $\beta\alp \simeq 0$ and $A_i$ equal to $-\infty$ outside the admissible radius and 0 inside.

\subsection{Mean-field expression for the search time}

A major asset of the foregoing generic framework is that it can be addressed not only by means of numerical simulations, but also more theoretically. Let us recall the major theoretical results that we obtained in this regard in \cite{dutta2022parking}, while referring the reader to that manuscript for the details of the derivation.

First, every street position associated with a parking spot as well as every intersection were handled as nodes of a `graph of spots' (this means that the street position where the car starts to park and the parking spot are amalgamated). This graph contains $N_{\mathrm{nodes}}$ nodes. 
The numbers of cars of category $\alpha$, i.e., $\alpha$-cars, passing
by each node per time unit is represented by a vector $\ul{I}\alp(t)$ of size $N_{\mathrm{nodes}}$, where $I_i\alp(t)$ is the rate of cars passing by node $i$ at time $t$, averaged over random realisations.
The drivers' turn choices at the nodes define a transition matrix $\uul{T}\alp$ such that $T_{ij}\alp\in[0,1]$ is the probability that an $\alpha$-car chooses to move from node $i$ to
node $j$ along an edge of the graph in one arbitrary time step, \emph{if} it does not park in the meantime. 
In this graph theoretical approach, $\alpha$-cars initially injected at nodes $j$ (hence, $I_j\alp(t=0)>0$) will be located
at positions represented by $\ul{I}\alp(t=1)= \ul{I}\alp(0) \cdot \uul{T}\alp$ at the next time step and at 
\begin{equation}
   \ul{I}\alp(K)= \ul{I}\alp(0) \cdot \Big(\uul{T}\alp\Big)^K 
\end{equation}
after $K$ steps, \emph{if they do not park in the mean-time}.
However, it is crucial to remark that cars may actually have parked in the meantime, with a probability $\tilde{p}\alp_i$ given (for each spot $i$) by  $\tilde{p}\alp_i=p_i\alp\,\hat{n}_i$, where $\hat{n}_i=1-n_i$ is zero (one) if the spot is vacant (occupied). Taking this possibility into account, the transition matrix $\uul{T}\alp$ should be substituted by $M\alp_{ij}= (1-p_i\alp\,\hat{n}_i)\cdot T_{ij}\alp$ and
the spatial distribution
of cars at $t=K$ is actually 
\begin{equation}
   \ul{I}\alp(K)= \ul{I}\alp(0) \cdot  \Big(\Ma\Big)^K.
\end{equation}

Provided that the occupancy field $(n_i)$ is known, the probability that an $\alpha$-car
reaches spot $j$ and parks there reads
\begin{equation}
    P\alp_{j} =  \underbrace{H_i\alp(0) \Big[\Big(\uul{\mathbb{I}} - \Ma\Big)^{-1}\Big]_{ij}}_{R_j\alp} \tilde{p}\alp_j, \label{eq:P_j_alp}
\end{equation}
where Einstein's summation convention (on repeated indices, excluding fixed index $j$ here) is implied, $\uul{\mathbb{I}}$ is the identity matrix, and $H\alp_j(0)= I\alp_j(0)/I\alp \in [0,1]$
is a renormalised rate, with $I\alp$
the total injection rate  of $\alpha$-cars. $R_j\alp$ denotes the probability to reach spot $j$ without accepting any parking spot before.

Along the same lines, the average `driving, searching, and parking' time $\Tk^{(\alpha,j)}$ of an $\alpha$-car finally parking at spot $j$ (in arbitrary time steps) can be derived; it is the average number of steps $K$ needed to park at spot $j$, weighted by the probability $H_j(K)\cdot \tilde{p}\alp_j$ to reach $j$ after $K$ steps and park there. Accordingly, summing over all spots $j$, and skipping the algebra detailed in \cite{dutta2022parking},
\begin{equation}
    \Tk^{(\alpha)}= H_i\alp(0) \cdot  \Big[ \Ma\cdot 
                     \Big(\uul{\mathbb{I}} - \Ma\Big)^{-2}\Big]_{ij}\cdot  \tilde{p}\alp_j.
\label{eq:stime_alpha_spots}
\end{equation}

\blue{
In reality, however, drivers will not keep cruising forever if they cannot find any vacant spot
and will quit searching for on-street parking after a given time, represented here by a maximum number of steps $K_{\max}$. Taking into account this upper bound, the foregoing expressions turn into 
\begin{equation}
   \bar{P}\alp_{j}= P\alp_j - H_i\alp(0)  \Big[\Big(\uul{\mathbb{I}} - \Ma\Big)^{-1} \cdot  \Ma^{K_{\max}+1} \Big]_{ij} \tilde{p}\alp_j
   \label{eq:P_j_alp_capped}
\end{equation}
\begin{equation}
\bar{\Tk}^{(\alpha)} = \Tk^{(\alpha)} - H_i\alp(0)\cdot\left[(\uul{\mathbb{I}}-\Ma)^{-2}\cdot \Ma^{K_{\max}+1}\right]_{ij} \tilde{p}\alp_j.
\label{eq:Ts_capped}
\end{equation}
Unfortunately, computing $\Ma^{K_{\max}+1}$ may be numerically very costly, as this is no longer a sparse matrix.
}

Before explaining how this complexity can be overcome, let us note that, in the above formulae, the search time was expressed in arbitrary units, each unit corresponding to a hop between two nodes of the `graph of spots'. Real time units can be recovered by making use of an auxiliary `generating' function
$\uul{N}(z)$ defined by $N_{ij}(z)=z^{\tau_{ij}}\,M_{ij}\alp$, where $z$ is a real variable and $\tau_{ij}$ is the travel time between neighbouring nodes $i$ and $j$ \cite{dutta2022parking}, viz.

\begin{equation}
T_{s}\alp =  H_i\alp(0) \cdot 
\left[(\uul{\mathbb{I}}-\Ma)^{-1}
\cdot N^\prime(z=1)
\cdot (\uul{\mathbb{I}}-\Ma)^{-1}
\right]_{ij}
\tilde{p}\alp_j,
\label{eq:stime_alpha_real}
\end{equation}
where the derivative of $\uul{N}(z)$ satisfies $N_{ij}^\prime(z=1)= \tau_{ij} M\alp_{ij}$

The foregoing formulae were derived for a \emph{given} configuration of the occupancy $\ul{n}$. To get the actual \emph{mean} search time requires averaging over an ensemble of equivalent realisations of $\ul{n}$. This step is tricky in general, but can be approximated by plainly substituting $\langle n_j \rangle \in [0,1]$ for $n_j =0$ or 1 in the definition of the $M_{ij}$ matrix (mean-field approximation).

\subsection{Stationary state occupancy}

Up to now, it has been assumed that the occupancy of each spot (or its time average) is known. This section explains how this occupancy field can be derived theoretically in the stationary regime. It is worth mentioning that the reasoning of \cite{dutta2022parking} is here extended to the important case of inhomogeneous departure rates $D\alp$.

This is achieved by writing a conservation equation, which balances incoming $\alpha$-cars and departing ones, viz.,
\begin{equation}
    \phi\alp = \frac{1}{N} \cdot  \frac{ I\alp}{D\alp},
\end{equation}
if all incoming drivers eventually manage to park.

In addition to this global balance, the rate at which $\alpha$-cars park \emph{at any given spot $j$} must be
balanced by the departure rate $D\alp$ of parked $\alpha$-cars, viz.,

\begin{equation}
      I\alp P_j\alp = D\alp \langle n_j\alp \rangle.
      \label{eq:balance_cat}
\end{equation}

It follows, using Eq.~\ref{eq:P_j_alp} and dropping the angular brackets, that $n_i\alp= \frac{I\alp}{D\alp}R_i\alp p_i\alp \hat{n}_i$ so that, summing over all categories $\alpha$, one finally arrives at 

\begin{equation}
    \hat{n}_i= \frac{1}{1+\sum_{\alpha}I\alp R_i\alp p_i\alp /D_i\alp},
    \label{eq:nj_stat_self}
\end{equation}
where $R_j\alp$, defined in Eq.~\ref{eq:P_j_alp}, implicitly depends on the $\langle n_i \rangle$'s.
This completes the derivation of the stationary occupation field $(n_i)$, insofar as Eq.~\ref{eq:nj_stat_self} is an implicit equation which self-consistently defines $(n_i)$ and can be solved by means of a fixed-point iterative method.

\subsection{Validation in a large-scale test case}
In Ref.~\cite{dutta2022parking}, we validated the theoretical approach on the large-scale
street network of the city of Lyon and
showed that  the foregoing formulae giving the stationary occupation field (Eq.~\ref{eq:nj_stat_self}) as well as the travel time by car category (Eq.~\ref{eq:stime_alpha_real}) are in excellent agreement with the steady-state results of numerical simulations of the agent-based model, for unbound search times. 
Unfortunately, the computational complexity of calculating $\Ma^{K_{\max}+1}$ in Eq.~\ref{eq:P_j_alp_capped}-\ref{eq:Ts_capped} hampered our endeavour to extend the comparison
to the more realistic case in which cars quit searching after a given time.

\section{Coarse-graining occupation fields at the street level}

To sum up, despite the success of the theoretical approach, there remains a difficulty associated with it: it involves
multiplications and inversions of matrices such as $\Ma$, with a linear size of order the number of spots in the network. This reflects the fact that parking decisions are made with respect to each parking space individually. The $\Ma$ matrices are particularly sparse and can therefore be handled with dedicated algorithms, but exponentiating these matrices is particularly inconvenient.

\subsection{Coarse-graining method}
Here, we aim to simplify the problem by coarse-graining the occupation fields at the level of the streets,
in order to be able to reason in terms of the `graph of streets', rather than the `graph of spots'. The
gist of this simplification consists in

(i) deriving an average occupancy $\phi_{\mathrm{street}}=\frac{1}{N_{\mathrm{s}}}\sum_{i=1}^{N_s} n_i$
per street link, where $N_s$ is the number of spots in the street, knowing the rate of incoming cars $I_{\mathrm{street}}$ and the characteristics of spots, and then 
 
 (ii) using these  $\phi_{\mathrm{street}}$ to define a coarse-grained counterpart to the $\Ma$ matrices of Eq.~\ref{eq:stime_alpha_real}.

 More concretely, for point (i), we take advantage of the fact that street links are linear, which enables us to derive the occupancies $n_i$ in a sequential way, starting with the first spot, $i=1$, etc. Let $I_{\mathrm{street}}\alp$ be the injection rate of $\alpha$-cars at the entrance of the street link. Then, applying Eq.~\ref{eq:balance_cat} to the first spot, 
 \begin{equation}
     \hat{n}_1= \frac{1}{1+ \sum_{\alpha} I_{\mathrm{street}}\alp p_i\alp/D\alp},
 \end{equation}
 which enables us to derive $R_2\alp= (1-\hat{n}_1 p_1\alp)$, and so on, until all $n_i$ have been calculated. (This operation takes a time proportional to the number of spots $N_s$.) Finally, the mean occupancy  $\phi_{\mathrm{street}}$ is obtained, and, along with it, the probability that an $\alpha$-car injected in the street exits from it without parking,
 \begin{equation}
     R_{\infty}\alp([I_{\mathrm{street}}\alp])= \prod_{i=1}^{N_s}(1-\hat{n}_i p_i\alp).
 \end{equation}
 
 To achieve point (ii), one simply has to notice that Eq.~\ref{eq:stime_alpha_real} still holds for the `graph of streets', provided that $\Ma$ is suitably adjusted. More precisely, in the coarse grained version, this matrix should turn into 
 \begin{equation}
 \Ma_{\mathcal{I} \mathcal{J}} \leftarrow [R_{\infty}\alp([I_{\mathrm{street}}\alp ])]_{\mathcal{I}}\,T\alp_{\mathcal{I} \mathcal{J}},
 \label{eq:Ma_new}
 \end{equation}
 where nodes $\mathcal{I}$ and $\mathcal{J}$ are now intersections marking the beginning of a street-link, and no longer spots. 
 
 The previous derivation within each street comes down to assuming that the spot occupancies within each street equilibrate (i.e., reach their
 stationary state) between every iteration of the fixed-point method for the whole network. While this may not be true from a dynamical perspective, it is reasonable to expect that it tends to the same fixed-point as the non-coarse-grained method.

\subsection{Validation}
At this stage, the coarse-grained method should be validated and its computational efficiency ought to be compared with that of the \emph{bona fide} method. Regarding the latter point, coarse-graining has reduced the linear size of the involved matrices $\Ma$ (in Eq.~\ref{eq:Ma_new}) by a factor of order $N_s$ (the number of spots per street), at the expense only of performing a number of order $N$ (the number of spots in the network) of operations at each iteration of the fixed point method. Accordingly, this strongly reduces the computational expense of all calculations involving these matrices, which are ubiquitous in the formulae we derived, and the reduction is all the stronger as street links contain many spots.

\begin{figure}[!htb]
    \centering
 
        \includegraphics[width=\textwidth]{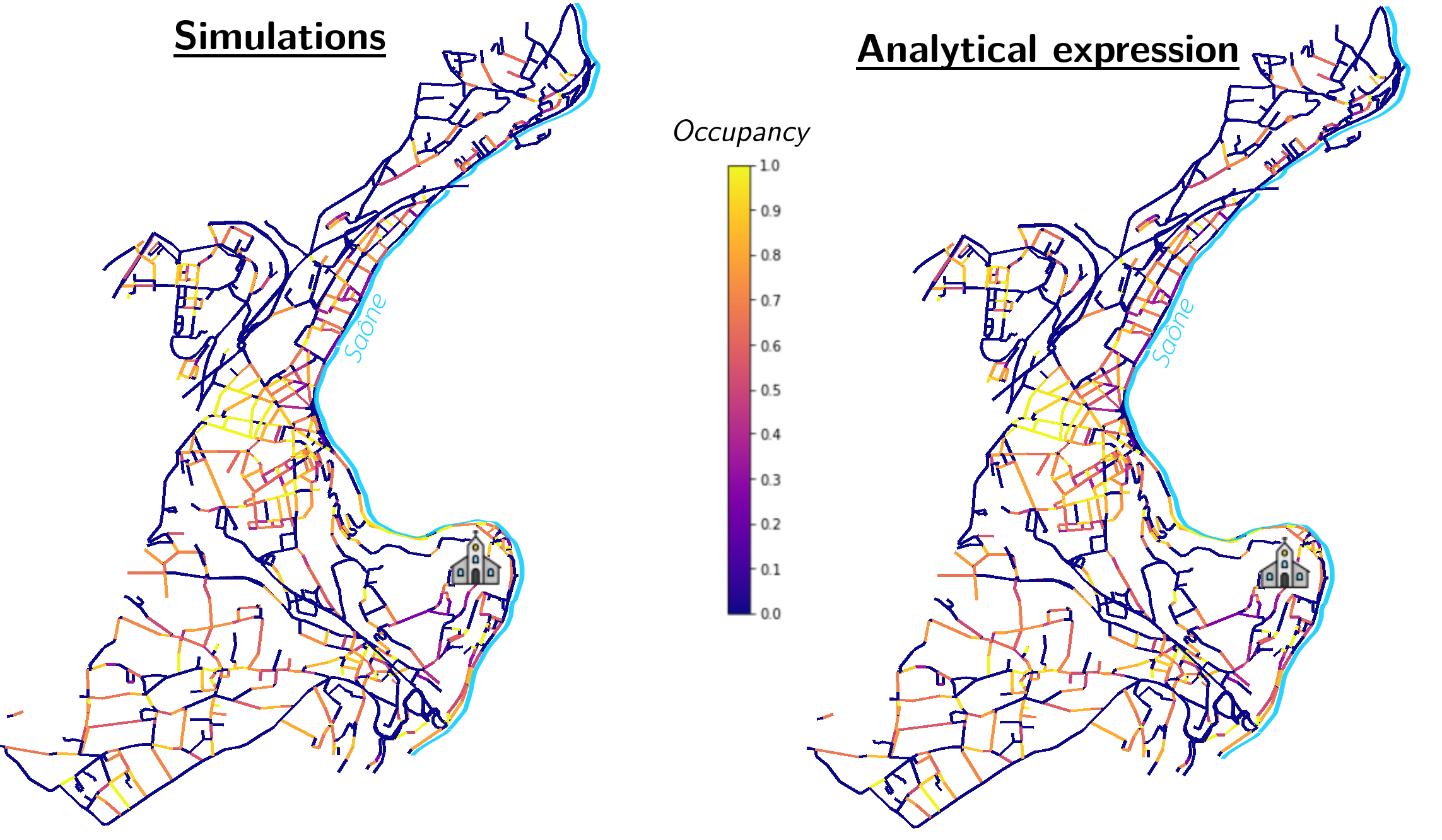}
\caption{Map of the stationary occupancy of street links in the Western part of Lyon (west of the river Sa{\^o}ne): Comparison of the  results obtained by numerical simulations of the agent-based model (left) and of the analytical predictions with the coarse-grained method described in this section (right). }
\label{fig:occ_small}
\end{figure}

\begin{figure}[!htb]
    \centering
        \includegraphics[width=0.65\textwidth]{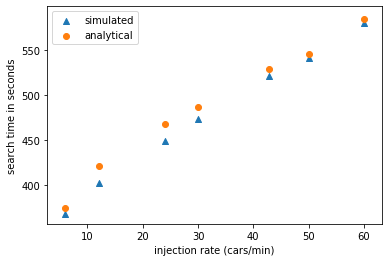}
\caption{Comparison of the search times obtained with direct numerical simulations and with
the coarse-grained analytical method introduced here. The maximal time before drivers quit searching was set to 15 minutes here. }
\label{fig:search_times_comp}
\end{figure}

\blue{
Turning to the validation, we considered the part of Lyon which lies to the West of the river Sa\^one, which represents about one third of the total street network. Eight car categories 
$\alpha=0 \dots 7$ are defined, each corresponding to a distinct destination within this zone, in line with what was done in \cite{dutta2022parking}; the turn-choice matrices $\Ta$ guide cars from their injection point to their destination along a route that is allowed to fluctuate around the shortest path, to some extent. For these simulations, the parameter $\beta$ controlling parking tension is set to 0.01 and 24 cars are injected per minute, while the mean parking duration is set to 20 minutes.
Most importantly, an upper bound was imposed on the cruising time: drivers quit searching for on-street parking after 25 minutes (we also tried 15 minutes). 
Previously, this capped condition could not be handled using our analytical formulae, because of the 
difficulty to exponentiate the per-spot matrix $\Ma$; this is now possible. Thus, an intractable equation has thus become within our reach. (On the other hand,
the convergence of our fixed-point method to determine the stationary occupancy may be tricky, which
has prevented us from theoretically handling the whole street network of Lyon so far.)
Figure~\ref{fig:occ_small} proves that the parking occupancy field obtained with our revised (coarse-grained)
analytical expressions are in very good agreement with the result of direct numerical simulations;
the slight differences (less than a few percent) mostly occur in high-density parking zones.
Furthermore, the travel times given by the analytical expressions also nicely reproduce the numerical outcome, as shown in Fig.~\ref{fig:search_times_comp} for various global car injection rates, for one category of drivers. The agreement is not quite as good as that found with the original method, for 
non-capped search times, prior to coarse-graining; this is not very surprising, insofar as considering the network at the level of street links instead of parking spaces introduces some inaccuracy in the assessment of the final driving time, in the last street-link.

}

\section{Conclusions}

In summary, we have built on a very recently proposed framework that generalises existing agent-based models for parking search and puts greater emphasis on factors such as the topology of the street network and the unequal attractiveness of parking spaces.  It was previously shown that, despite its generality, this model can be solved analytically in the mean-field stationary regime. A matricial formula relating the total driving time (including the search time) to the occupancy of parking spaces was thus derived. Here, the formula was extended to allow different categories of drivers to have different parked times (i.e., departure rates), which is naturally of practical relevance.

Furthermore, the foregoing formula was fairly cumbersome, involving very large matrices. In this contribution, we have demonstrated that the analytic expression can be be further simplified by aggregating parking spots by street link, so that one now handles a `graph of streets' instead of a `graph of spots' (the former containing much fewer nodes, of course). This simplification drastically reduces the dimension of the matrices involved in the foregoing formula and makes them even more tractable, which could be used efficiently by transport engineers. It paves the way for a treatment of practical issues which would otherwise be computationally costly to simulate, in particular optimisation problems in the context of redesigns of the transport network.

For sure, our model currently presents some limitations, notably the lack of interactions between the cruising traffic and the underlying one, as well as the absence of feedback between the experienced search times and the parking demand. There is no reason why these limitations could not be overcome in the near future; for instance, the second limit can be overcome by integrating our model
for parking search as a module in a multimodal choice model.

\section*{Acknowledgments}
We acknowledge funding from IDEXLYON and Institut Rh\^onalpin des Syst\`emes Complexes (IXXI).

\end{document}